\documentclass[aps,prx,twocolumn,superscriptaddress,longbibliography]{revtex4-1}
\usepackage{graphicx}

\usepackage{xcolor}

\begin{document}
\title{Diverse strategic identities induce dynamical states in evolutionary games}

\author{I. Sendi\~na-Nadal*}
\affiliation{Complex Systems Group \& GISC,  Universidad Rey Juan Carlos, 28933 M\'ostoles, Madrid, Spain}
\affiliation{Center for Biomedical Technology, Universidad Polit\'ecnica
de Madrid, 28223 Pozuelo de Alarc\'on, Madrid, Spain}
\email[Corresponding author: ]{irene.sendina@urjc.es}

\author{I. Leyva}
\affiliation{Complex Systems Group \& GISC,  Universidad Rey Juan Carlos, 28933 M\'ostoles, Madrid, Spain}
\affiliation{Center for Biomedical Technology, Universidad Polit\'ecnica
de Madrid, 28223 Pozuelo de Alarc\'on, Madrid, Spain}

\author{M. Perc}
\affiliation{Faculty of Natural Sciences and Mathematics, University of Maribor, Koro{\v s}ka cesta 160, 2000 Maribor, Slovenia}
\affiliation{Department of Medical Research, China Medical University Hospital, China Medical University, Taichung, Taiwan}
\affiliation{Complexity Science Hub Vienna, Josefst{\"a}dterstra{\ss}e 39, 1080 Vienna, Austria}

\author{D. Papo}
\affiliation{SCALab, CNRS, Universit\'e de Lille, Villeneuve d’Ascq, France}

\author{M. Jusup}
\affiliation{Tokyo Tech World Research Hub Initiative (WRHI), Institute of Innovative Research, Tokyo Institute of Technology, Tokyo 152-8552, Japan}

\author{Z. Wang}
\affiliation{School of Mechanical Engineering, Northwestern Polytechnical University, Xi'an 710072, China}

\author{J.A. Almendral}
\affiliation{Complex Systems Group \& GISC,  Universidad Rey Juan Carlos, 28933 M\'ostoles, Madrid, Spain}
\affiliation{Center for Biomedical Technology, Universidad Polit\'ecnica
de Madrid, 28223 Pozuelo de Alarc\'on, Madrid, Spain}

\author{P. Manshour}
\affiliation{Physics Department, Persian Gulf University, Bushehr 75169, Iran}

\author{S. Boccaletti}
\affiliation{CNR, Institute of Complex Systems, Via Madonna del Piano 10, 50019 Florence, Italy}
\affiliation{Unmanned Systems Research Institute, Northwestern Polytechnical University, Xi'an 710072, China}

\begin{abstract}
Evolutionary games provide the theoretical backbone for many aspects of our social life: from cooperation to crime, from climate inaction to imperfect vaccination and epidemic spreading, from antibiotics overuse to biodiversity preservation. An important, and so far overlooked, aspect of reality is the diverse strategic identities of individuals. While applying the same strategy to all interaction partners may be an acceptable assumption { for  simpler
forms of life, this fails to account} for the behavior of more complex living beings. For instance, we humans act differently around different people.
Here we show that allowing individuals to adopt different strategies with different partners yields a very rich evolutionary dynamics, including time-dependent coexistence of cooperation and defection, system-wide shifts in the dominant
strategy, and maturation in individual choices. Our results are robust to variations in network type and size, and strategy updating rules. Accounting for diverse strategic identities thus has far-reaching implications in the mathematical modeling of social games.
\end{abstract}

\maketitle

Game theory owes its popularity to its ability of describing social interactions as well as the gist of conflicting situations in economics and politics \cite{neumann_44}. An important paradigm shift was the introduction of evolutionary games \cite{maynard_82}, which focus on the interaction and competition between different strategies. Today, evolutionary games are the workhorse for studying frequency dependent selection in biological system, and they are used prolifically to investigate cooperation and competition in economic, social, and technological systems \cite{nowak_06, javarone2018book}. In general, evolutionary games allow modelers to make the interactions within a system measurable and interpretable, which pairs particularly well with network science {\cite{albert_rmp02, boccaletti_pr06, boccaletti_pr14, kivela_jcn14,Fu2017,Khoo2018}}.

A frequently used subclass of evolutionary games is that of social dilemmas, in which what is best for the individual player conflicts with what is best for the society as a whole \cite{nowak_s06}. Such situations are commonplace in modern human societies, applying as much to the abuse of power in leadership as to climate inaction and the over-exploitation of natural resources. Much research has been devoted to the resolution of social dilemmas on networks \cite{ohtsuki_n06, santos_pnas06, santos_n08, tanimoto_pre07, fu_pre09, lee_s_prl11, fu2017leveraging, allen2017evolutionary, fotouhi_rsif19}, with results pointing to the fact that the network topology (and its possible temporal changes) is essential to determine the system's evolution \cite{perc_bs10, perc_jrsi13}. For instance, hubs in scale-free networks can act as strong facilitators of cooperation \cite{santos_prl05}, extending the survival range well past the boundaries of traditional network reciprocity \cite{nowak_n92b}. Moreover, a solution for weak selection applicable to any kind of graph has been recently obtained based on calculating the coalescence times of random walks \cite{allen2017evolutionary}.

\begin{figure*}
\includegraphics[width=0.75\textwidth]{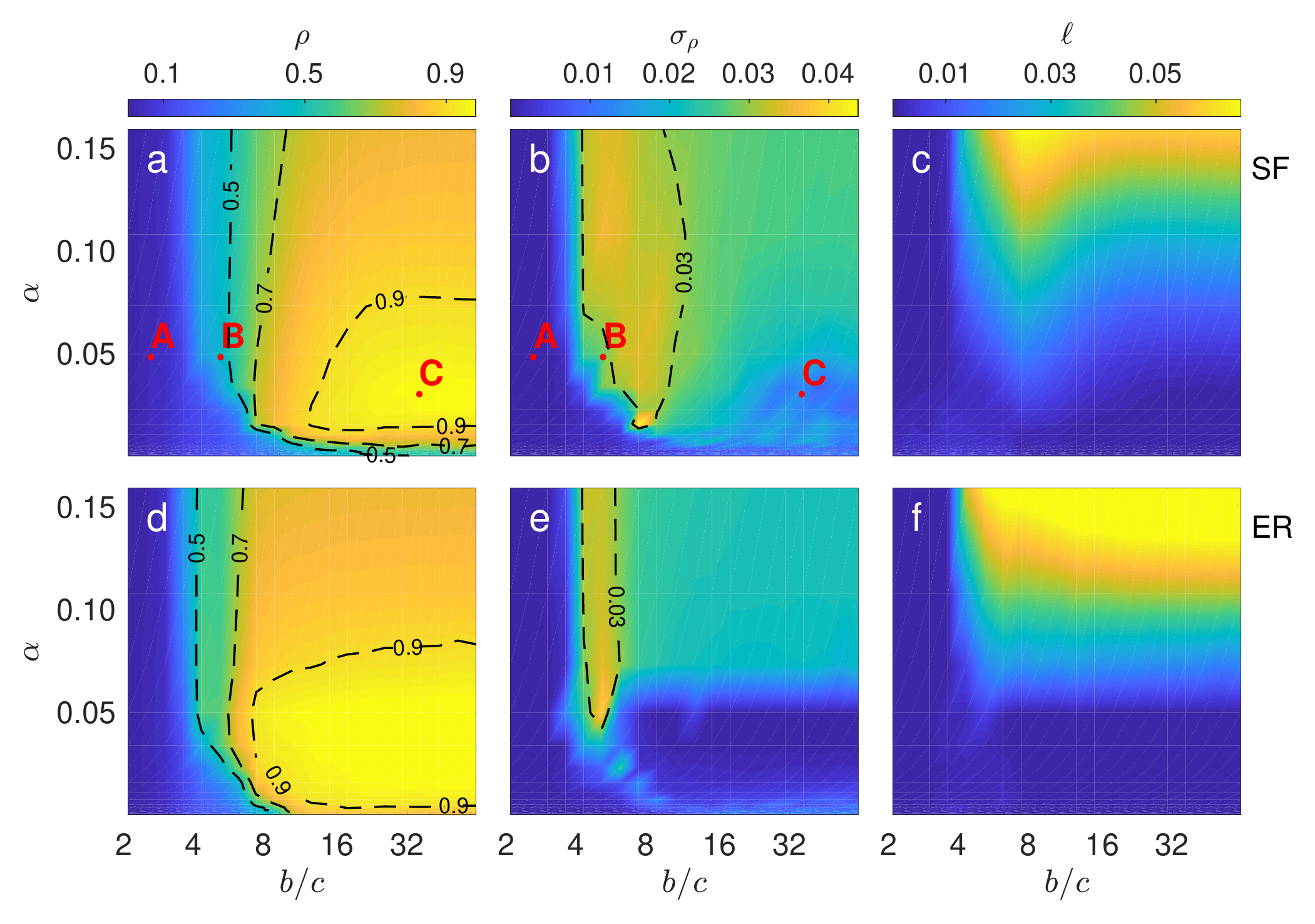}
\caption{{\bf Diversity in strategic identities induces time-dependent states}. Heatmaps (with colorbar on top) of $\rho$ (panels a and d), $\sigma_\rho$ (panels b and e), and  $\ell$ (panels c and f) in the parameter space $(\alpha,b/c)$.
See text for the definitions of all quantities. Panels a-c refer to SF networks, whereas panels d-f refer to ER graphs. Simulations were performed over $T_f = 10,000$ times units, while the time-averaging of all quantities was made over the last $T=2,000$ steps. Each point is obtained by a further ensemble average over 10 network realizations with size $N = 1,000$ and average degree $ \langle k\rangle=4$. Isolines at the values of $\rho=$0.5, 0.7, and 0.9 are drawn in panels a and d. The isoline at the value of $\sigma_\rho=$0.03 is reported in panels b and e.
Three different regimes can be identified: i) a stationary state in which cooperation dominates the network, at sufficiently large $b/c$ ratios and sufficiently small $\alpha$ values (the red point labeled with C and located at (36,0.03) in panels a,b); ii) the classical regime for $b/c<  \langle k\rangle$ where defection becomes dominant and cooperators can no longer survive (the red point labeled with A and located at (2.5,0.048) in the same panels); and iii) a novel {\it dynamical state} with coexisting cooperators and defectors (the red point labeled with B and located at (5,0.048)). The emergence of the new state {\it is a direct consequence} of diversity in the players' strategic identities.}
\label{fig1}
\end{figure*}

So far, studies of evolutionary dynamics in structured populations have typically assumed that each individual adopts a single strategy with all their interaction partners \cite{szabo_pr07}.
For organisms with no or limited self-awareness and intelligence, this assumption { may be indeed} reasonable:
{ bacteria, for instance, are not always able to change their genotype} based on their surroundings \cite{stefanic2015kin}, and similarly a worker bee can hardly break with the hive and go about exploring alternatives.
However, the assumption becomes unrealistic for more complex living beings.
Because evolutionary games are increasingly used to study complex phenomena in human societies \cite{rand_tcs13, vasconcelos_ncc13, pacheco_plrev14, szolnoki_jrsif14, pastor_rmp15, wang_z_pr16, perc_pr17, chen_fp18, chen_prsb19}, the time is ripe -- actually the need is urgent -- for the relaxation of this monotony, and for the account of the obvious fact that we act differently with certain people than we do around others \cite{marcia1966development, erikson1994identity}. And this is of course likewise true for many other animals.

Moreover, monotonic strategies lead, in general, to time stationary network's arrangements, implying the setting (from a given time on) of a population of {\it simpletons} where the identity of each unit is that of a permanent cooperator or of a permanent defector. This is also far from properly representing real interactions in human or animal societies, wherein members actually alternate cooperation and defection in time. In social systems, for instance, it has been observed that human propensity to cooperation does vary in time, and this has important consequences for the outcomes in decision conflicts \cite{Evans2015}, competitive environments \cite{Cone2014}, and for long-run cooperation \cite{Mao2017}. In biophysical systems, changes in cooperative behavior are observed over time due to population feedbacks \cite{Sanchez2013} or to varying resource availability \cite{Hoek2016}. Varying cooperation propensities characterize, for instance, microbial populations, where microbes are known for their dualistic roles, i.e., cooperate with  some strains but defect with others \cite{Celiker2013}.

We then here abandon the monotonic assumption, and show that taking instead into explicit account diverse players' identities induces a very rich
dynamical repertoire, which includes several of such observed phenomena: time-dependent coexistence of cooperation and defection, systemic shifts in the network's dominant strategy, and the maturation of individual strategic choices. Overall, our results show that it is crucial for individuals to carefully gauge their strategic vectors for socially optimal evolutionary outcomes, thus revealing that diverse strategic identities are essential for properly capturing a vast number of behaviors observed in human societies.

We start by considering a network, $G$, made of $N$ players (or units, or nodes), connected according to a symmetric adjacency matrix, $A=(a_{ij})$, where $a_{ij}=1$ if units $i$ and $j$ interact, and $a_{ij}=0$ otherwise. Moreover, $k_i=\sum_{j=1}^N a_{ij}$ is the number of neighbors of player $i$, also called node $i$'s degree. A strategy matrix, $S=(s_{ij})$, is introduced such that $s_{ij}=1$ if player $i$ cooperates with player $j$, $s_{ij}=2$ if player $i$ defects from $j$, and $s_{ij}=0$ if $a_{ij}=0$.

\begin{figure*}
\includegraphics[width=0.85\textwidth]{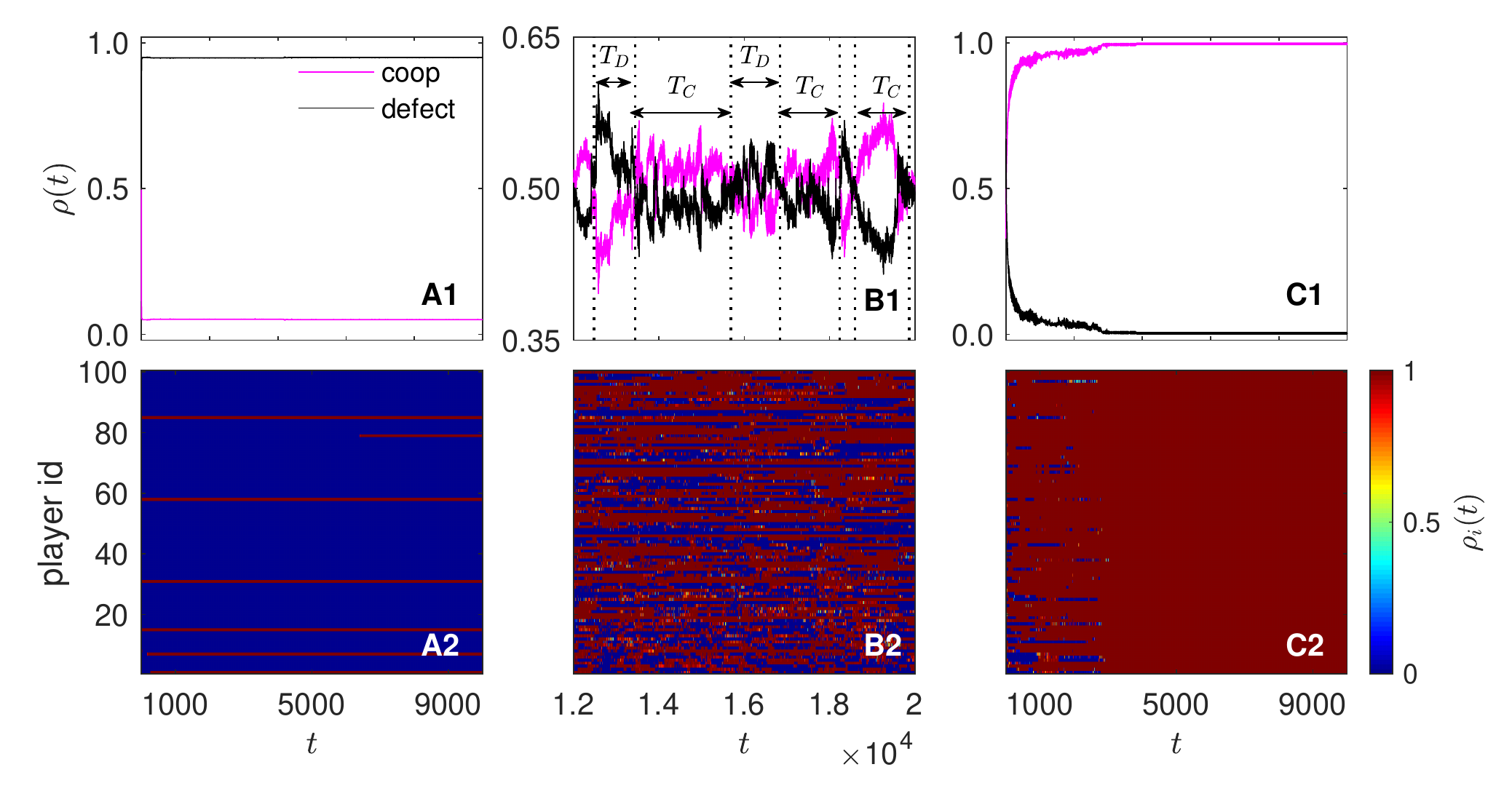}
\caption{{\bf The coexistence of cooperators and defectors is characterized by system-wide shifts in the network dominant strategy.} Time evolution of $\rho(t)$ (first row) and of the cooperation level $\rho_i(t)$ of the 100 larger-degree players in the network (second row, with color bar at the right). See text for definition of all quantities. Data refer to a single run of the game on a single realization of an SF network with $N=1,000$ and $ \langle k  \rangle=4$. The game parameters correspond to those marked by red points A (first column), B (second column), and C (third column) in Fig.~\ref{fig1}a). In panel A1, defection dominates over cooperation in a steady state, and all the strategies of individual players become soon constant in time (panel A2). Conversely, in panel C1 cooperators ultimately dominate, though temporal fluctuations persist over a relatively large time scale. The corresponding temporal profile of individual strategies (panel C2) exhibits fast temporal changes during the early stages of the game before maturation of the individual strategic choices.
In the novel state (panels B1 and B2), the coexistence of cooperators and defectors is characterized by system-wide shifts in the dominant strategy. The horizontal lines in panel B1 mark the transitions from periods during which cooperation is dominant (intervals with label $T_C$) to time lapses during which defection dominates instead (intervals with label $T_D$). Maturity never takes place here, and players are forever trapped in cycles of individual changes (panel B2).}
\label{fig2}
\end{figure*}

{Our approach is in line with a stream of recent research where edges (rather than vertices) are the centers of evolutionary dynamics \cite{Wardil2009,Wardil2010,Wardil2011,Su2016,Su2019}. In particular, Ref.~\cite{Wardil2009} considered the ability to change the strategy along the least productive edge (in terms of payoff) as a form of punishment that ultimately promotes cooperation, and the result has been later extended from rings and lattices to well-mixed populations \cite{Wardil2010} and eventually to heterogeneous networks \cite{Wardil2011}. On the other hand, Ref.~\cite{Su2016} assumes that the edge strategy (cooperation or defection) yielding a higher payoff is likely to replace strategies producing lower payoffs, and shows that this mechanism generally supports cooperation. { Ref.~\cite{Su2019} extends the idea of edges being the centers of evolutionary dynamics by suggesting that edges may also convey additional information, such as genetic similarity, geographic proximity, and social closeness. Moreover, Ref. \cite{Javarone2017} shows how the heterogeneity of a group, in terms of skills and strategies, can have implications at evolutionary level in the stability of communities and in their size. Finally, in Ref. \cite{Javarone2017b} the public goods game is used to solve the Traveling Salesman Problem, under the hypothesis that agents may play with a vector of strategies (each one corresponding to a possible solution of the game), and therefore diversification is somehow taken into account.}
We here show for the first time that allowing individuals to adopt different strategies with different partners yields a very rich evolutionary dynamics, including time-dependent coexistence of cooperation and defection, system-wide shifts in the dominant
strategy, and maturation in individual choices.}
{ Other approaches considered situations where a game with one person can affect the results of a game with another person (see, for instance, Ref.~\cite{Reiter2018}, where the authors refer to this phenomenon as {\em crosstalk}. An implicit assumption needed to implement crosstalk is that an individual can distinguish between other individuals with whom they play, which is an explicit assumption in our study. Crosstalk impedes cooperation and requires more forgiving strategies than, say, tit-for-tat, for cooperation to flourish.}

Each player $i$ is then here associated to a {\it vector of independent strategies}, the dimension of which is degree $k_i$. In other words, in a particular instance of a game, players can simultaneously cooperate with some of their neighbors and defect with others. One further has $k_i=k_i^C+k_i^D$, where $k_i^C=\sum\limits_{j\vert s_{ij}=1} a_{ij}$ ($k_i^D=\sum\limits_{j\vert s_{ij}=2} a_{ij}$) is the time-dependent number of neighbors player $i$ is currently cooperating (defecting) with.

In the most representative and widely applicable social dilemma, the prisoner's dilemma \cite{rapoport1965prisoner}, the interactions between player pairs are governed by payoff matrix
\[
   P=\,
\begin{array}{c|cc}
   & C & D \\
  \hline
  C & b-c & -c \\
  D & b & 0
 \end{array}
\]
which means that if player $i$ cooperates and player $j$ defects ($C\leftrightarrow D$), then player $i$ incurs cost $-c$ while player $j$ collects benefit $b$, and vice versa. On the other hand, when two players cooperate among themselves ($C \leftrightarrow C$) they both acquire amount $b-c$, corresponding to the collected benefit from cooperation of the other agent minus the paid cost to cooperate. If two agents mutually defect ($D\leftrightarrow D$), they both obtain a zero payoff. When $b>c>0$, the Nash equilibrium is mutual defection \cite{nash_pnas50}.

The gain or loss, $g_{ij}$, along each existing ($a_{ij}=1$) link is calculated using payoff matrix $P$ as $g_{ij}=P(s_{ij},s_{ji})$.
The average payoff of player $i$ is therefore
\begin{equation}
g_i=\frac{1}{k_i}\sum_{j\in\Lambda_i} g_{ij},
\end{equation}
where $\Lambda_i$ is the set of neighbors of $i$. One can further calculate each agent's average payoff from being a cooperator or a defector separately, according to $g_i^C=\displaystyle\frac{1}{k_i^C}\sum_{j\in \Lambda_i \vert s_{ij}=1} P(C,s_{ji})$ and $g_i^D=\frac{1}{k_i^D}\sum_{j\in \Lambda_i \vert s_{ij}=2} P(D,s_{ji})$, respectively.
At time $t$, the network state can be tracked at various scales: microscopically, one monitors each player's cooperation level $\rho_i(t)=\frac{k_i^C(t)}{k_i}$, while macroscopically one refers to the network average $\rho(t)=\frac{\sum_i \rho_i(t)}{N}$. Moreover, $ \rho  = \langle \rho(t) \rangle _T$ and and $\sigma_\rho$ are, respectively, the time-averaged cooperation density over an observation time $T$ and the standard deviation of $\rho(t)$.

Each player $i$ is assumed to imitate the strategy adopted against it by its neighbor with the highest total payoff. Namely, at each step of the game, after players have collected their payoffs according to their current strategies' vectors, each player $i$ considers
\begin{equation}
\label{deltag}
\Delta g_{i}= \frac{g_i^{\tilde{s}}-g_i}{b},\label{game2}
\end{equation}
where $g_i^{\tilde{s}}=\max\limits_{j\in\Lambda_i}(g_{ji})$ is the payoff of the neighbor $j={\tilde{s}}$ inside $\Lambda_i$ who earned the maximum.

\begin{figure*}
\includegraphics[width=0.75\textwidth]{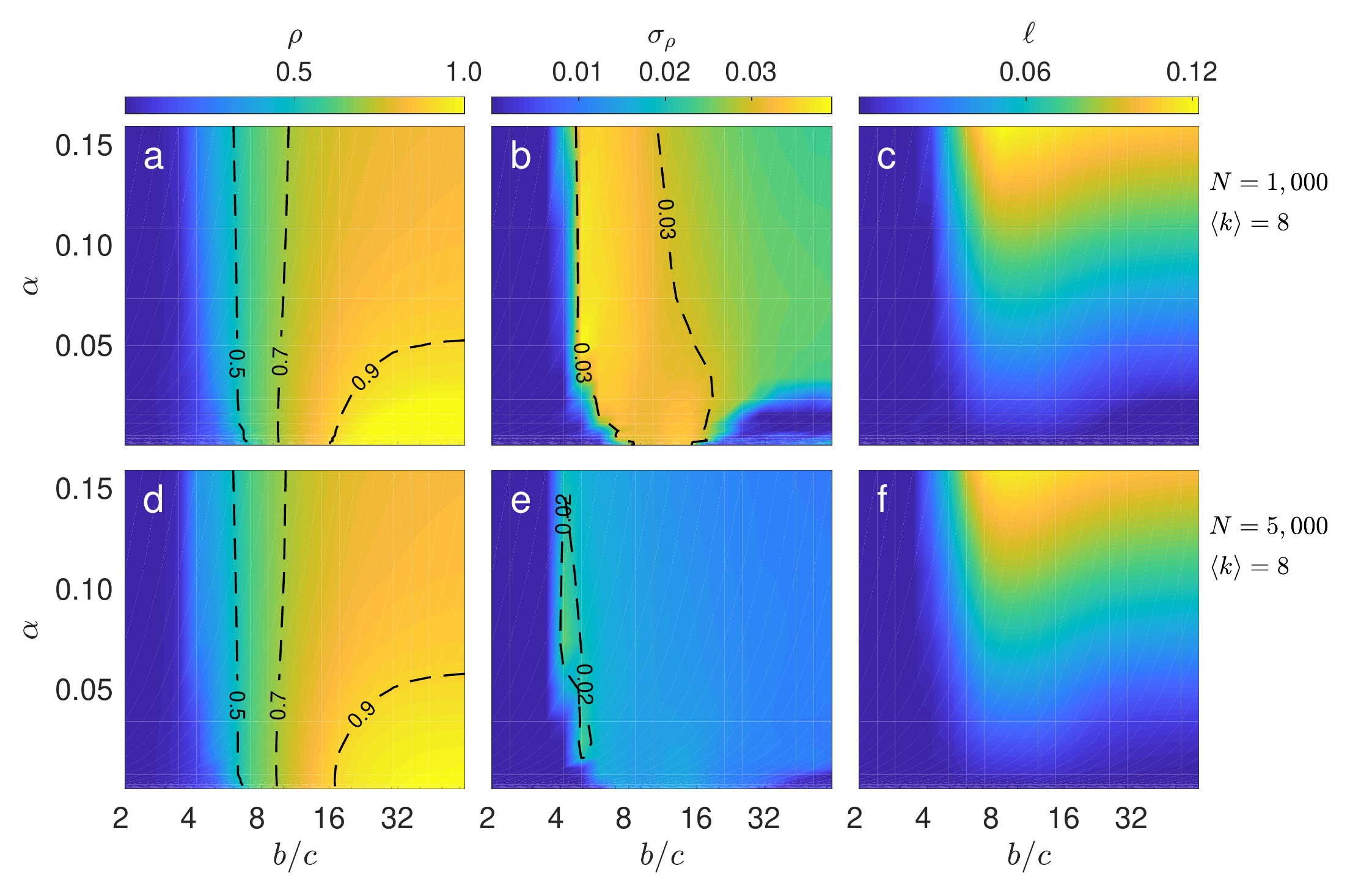}
\caption{{\bf  Increasing the connectivity density enhances time fluctuations, and the overall scenario is preserved with changing network size.} Heatmaps (colorbars on top) of $\rho$ (panels a and d),  $\sigma_\rho$ (panels b and e), and $\ell$ (panels c and f) in the parameter space $(\alpha, b/c)$. As compared with Fig.~\ref{fig1}, data here refer to SF networks with a larger average degree (top row, $N=1,000, \   \langle k  \rangle=8$), and with a larger size (bottom row, $N=5,000, \   \langle k  \rangle=8$). Each point is furthermore obtained by an ensemble average over 10 different network realizations. For a better visualization, isolines are drawn in panels a, b, d, and e. The main effect of a larger $ \langle k  \rangle$ is that of enhancing the fluctuations in the time-dependent state, as it is visible on panel b). On the other hand, the network size has the effect of slightly reducing the size of fluctuations (panel e), yet it has virtually no impact on the overall scenario (panels a and d), as well as on the diversity of strategic identities (panels c and f).}
\label{fig3}
\end{figure*}

Then, the player updates its $k_i$ strategies, i.e., it imitates the strategy ($C$ or $D$) used against it by its best neighbor along each one of its $k_i$ links with probability
\begin{equation}
\label{fermifunction}
\displaystyle p=\frac{1}{1+e^{-\frac{\Delta g_i}{\alpha}}},
\end{equation}
where $\alpha$ is a free parameter playing the role of an effective temperature in the above Fermi-Dirac function, while $\Delta g_i$ is the payoff balance (in units of the benefit $b$) given by Eq.~(\ref{game2}). If player $i$ was already adopting the best neighbor's strategy along a given link $ij$, it simply keeps that link unchanged.
According to these definitions, it may also happen that a given node $i$ imitates, e.g., the $C$ strategy because that is what its best neighbor (node ${\tilde{s}}$) played against it even if $g_{{\tilde{s}}}^C < g_{{\tilde{s}}}^D$. Put alternatively, the player adopts $C$ by imitating its best neighbor although that neighbor is earning more from $D$ than $C$.

We begin describing the results by showing in Fig.~\ref{fig1} three color maps that, from left to right, encode (i) the cooperation density $\rho$, (ii) the associated standard deviation $\sigma_\rho$, and (iii) the link strategy index defined by $\ell_i=1-\frac{|{k_i^D-k_i^C}|}{k_i}$. This latter index quantifies the {\it rigidity} of the strategy vectors: the index vanishes when players have strategy vectors made of all cooperation or defection entries, whereas it gets larger and larger the more diversified the players' identities are.
The color maps are represented in the parameter plane of the system, which is made of the cost-to-benefit ratio, $b/c$, and of temperature $\alpha$ in the Fermi-Dirac function. Moreover, the upper row shows the results obtained on scale-free (SF) networks \cite{barabasi_s99}, while the bottom row shows the results obtained on random Erd\H{o}s-R{\'e}nyi (ER) graphs \cite{erdos_pmd59}. To make the content of the Figure easily understandable, isolines are drawn in panels a, b, d, and e.

In structured populations, natural selection favors cooperation if the benefit of the altruistic act $b$ divided by cost $c$ exceeds the average degree of the network $\langle k \rangle$ \cite{ohtsuki_n06}. In our case, cooperation can evolve as a consequence of social viscosity even in the absence of reputation effects or strategic complexity. The $b/c > \langle k \rangle$ rule is therefore a good reference against which we can compare the impact of diverse strategic identities on the evolution of cooperation.

Panels a and d of Fig.~\ref{fig1} reveal that cooperation dominates the entire network for sufficiently large $b/c$ ratios and sufficiently small $\alpha$ values. This regime is marked with the red point C. When decreasing $b/c$, a rather clear transition occurs at $b/c \sim \langle k \rangle$ toward the classical regime marked with the red point A in the same panels, where defection becomes dominant and cooperators no longer survive. In between these two extremes, however, a novel {\it dynamical state} emerges. The new state consists of coexisting cooperators and defectors, and is maintained by the evolutionary dynamics, yet in a distinct and time-dependent way. Namely, by looking at the ensemble mean of the standard deviation and the link strategy index shown in the middle and rightmost panels, respectively, one can observe that different levels of cooperation are associated with different evolutionary dynamics: when cooperators and defectors coexist, this is characterized by a large standard deviation and also by larger values of the link strategy index, as indicated by the red point B in panels a and b. Conversely, when either cooperators or defectors dominate, the standard deviation is lowest, as is the link strategy index. The new state, therefore, {\it is a direct consequence} of the diversity in the players' strategic identities; when diversity is lost and one recovers the case of a single strategy per player, only stationary states are observed and for sufficiently large $b/c$ these are made of all cooperators, or for $b/c<\langle k \rangle$ of all defectors.

Furthermore, comparing the two rows in Fig.~\ref{fig1} indicates that, while our observations are equally valid for
SF and ER networks and are thus robust against variations in topology, temporal fluctuations are stronger in
heterogeneous wirings. This is relevant because interactions in human societies are indeed characterized by not
trivial, yet heterogeneous, connectivity patterns \cite{Watts1998,Watts1999,Newman2001,Gallos2013,pastor_rmp15,Broido2019}.

Figure~\ref{fig2} reveals the details behind the dynamics emerging at the three points marked by letters A, B, and C in Fig.~\ref{fig1}a. Looking at case A, one can observe an almost fully steady time course of both strategy densities (panel A1). Eventually, defection dominates over cooperation, and the strategies of the individual player do not change in time (panel A2 containing the identity of the 100 most connected nodes in the network).

Case C, in which cooperators dominate after a transitory dynamics, exhibits a somewhat different evolution from case A. Already early in the evolutionary process cooperators seem destined for dominance, but residual fluctuations persist over a relatively large time scale, before eventually the population reaches the absorbing steady state (panel C1). This is corroborated by the temporal profile of individual strategies (panels C2), which exhibits fast temporal changes during the early stages of the evolutionary process before defection recedes completely. One here observes a maturation in the strategic choices of individuals across their whole vector of strategies.

Case B is of particular interest as it is characterized by the largest ensemble mean of the standard deviation. Here, the coexistence of cooperators and defectors is characterized by system-wide shifts in the dominant strategy (panel B1). At odds with case C, maturity never takes place, and players are forever trapped in a cycle of individual changes that allows time-dependent cooperation, but never full dominance (panel B2). In this case, diverse strategic identities are thus a successful vehicle for maintaining the survival of cooperation even at strongly unfavorable $b/c$ ratios, but not for assuring the elimination of defectors.

{ We have checked that regime C (full cooperation) and regime A (full defection) are always (i.e. regardless on the specific network realization) attained by the system  as long as the initial number of cooperating ties is, respectively, above and below 10 \%. As for point B, we have verified that temporal fluctuations are really persistent in the system, as they were supported over time scales of the order of 1 million iterations of the game.}

\begin{figure*}
\includegraphics[width=\textwidth]{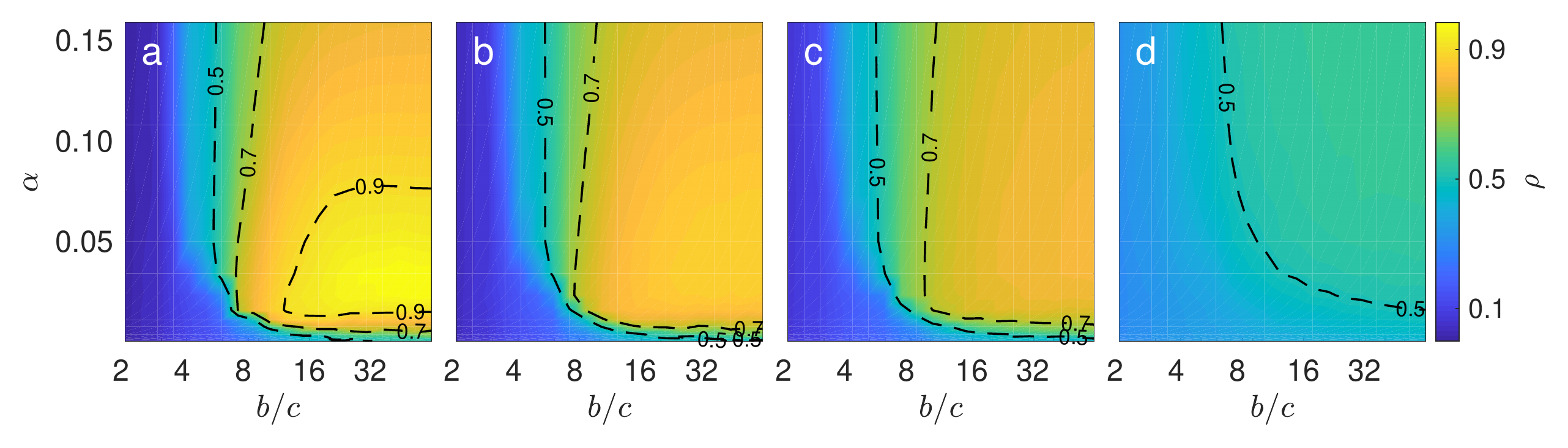}
\caption{{\bf Diversity and time-dependent states are observable also for more realistic, heterogeneous, strategy rules}. Heatmaps (colorbar on the right) of the cooperation density $\rho$ in the parameter space $(\alpha, b/c)$. Here, the network is partitioned in two communities: a fraction $p$ of randomly selected players adopts an introspective strategy updating (see text for details), whereas the remaining fraction $1-p$ of players adopt the usual imitation rule. Panels a-d correspond, respectively, to $p = 0$, $p=0.05$,  $p=0.1$, and $p = 0.5$. As in Fig.~\ref{fig1}, each point is obtained by an ensemble average over 10 different realizations of SF networks with $N = 1,000$ and $\langle k \rangle=4$. Isolines at $\rho=$ 0.5, 0.7, and 0.9 are drawn. Other stipulations as in the caption of Fig.~\ref{fig1}.
The results point to a remarkable robustness of the observed scenario also in the presence of heterogeneity in the strategic updating rules, and therefore corroborate the fact that diverse strategic identities can be observed in a plethora of practical cases.}
\label{fig4}
\end{figure*}

{  It has to be noticed, at this stage, that the dynamical scenario reported in Figs. 1 and 2 is obtained under the assumption that the strategy a given node $i$ is
selecting at each step against each one of its neighbors in $\Lambda_i$ depends only indirectly on the payoffs collected by those nodes in the previous step. Indeed, while the strategy taken as reference for imitation is that played by the neighbor with the largest accumulated payoff (and therefore its determination includes a comparison among all payoffs of the neighbors of node $i$), the argument of the Fermi-Dirac function in Eq. (\ref{fermifunction}) is equal for all the nodes in $\Lambda_i$. However, we have verified that exactly the same scenario (with only suitable rearrangements of the parameters $\alpha$ and $b/c$) occurs also in the case in which node $i$, once adopted the reference strategy as that used against it by its best neighbor, imitates such a strategy with different probabilities in different links, namely by calculating in each link $ij$ the Fermi-Dirac function with reference to the difference between the payoff of node $j \in \Lambda_i$ and the payoff of node $i$. Our choice is therefore dictated by practical reasons: instead of having to have the exact information on all payoffs accumulated by its neighbors, we suppose that each node $i$ needs only to know the payoff of the best performing node in its neighborhood $\Lambda_i$ and the strategy that such a node played against it in the previous step.

We also tested the robustness of the results against changes in network size and connectivity density. We} show in Fig.~\ref{fig3} the same set of color maps as in Fig.~\ref{fig1}, but reporting here the results obtained with SF networks with a larger average degree (Fig.~\ref{fig3}a-c, $N=1,000, \ \langle k  \rangle=8$), and with a larger size (Fig.~\ref{fig3}d-f, $N=5,000, \  \langle k  \rangle=8$). Contrasting the top row of Fig.~\ref{fig1} with Fig.~\ref{fig3}, one realizes that the most fascinating features are all maintained.
The data are slightly shifted to higher $b/c$ ratios, but not as much as they should have been if they would have strictly followed the $b/c> \langle k\rangle$ rule. This is not surprising given that the derivation of the rule was made in Ref. \cite{ohtsuki_n06} for regular networks, and in the limit of very sparse graphs ($N \gg  \langle k  \rangle$), and therefore it has to be considered as a necessary, but not sufficient condition for the transition to full cooperation.

Examining the details of Fig.~\ref{fig3}, one also realizes that the main effect of a larger $ \langle k  \rangle$ is that of enhancing the fluctuations in the time-dependent state (panel b). { We also performed simulations (not shown here) at lower ($\langle k  \rangle=2$) and higher ($\langle k  \rangle=12$) values of the average degree, which confirm the enhancement of fluctuations with increasing $\langle k  \rangle$.}
The network size has instead the effect of slightly reducing the size of fluctuations, yet it has virtually no impact on the overall scenario as well as on the diversity of strategic identities. Our results are then validated not just in terms of variation in network topology, but also in terms of network size and average degree. We have also tested our results on smaller size networks (with only $N=100$ players), and we observed identical outcomes. In terms of social models, diverse strategic identities thus yield consistently similar results for networks comprising just tens or thousands of individuals, and are therefore applicable for small groups and communities, upwards to big societies.

Notice that we performed simulations also in the limit of weak selection ($\alpha > 1$), which confirmed (up to $\alpha > 10$) the general scenario of a transition between a fully defective (at low values of $b/c$) and a fully cooperative (at high values of $b/c$) state mediated by a time-dependent configuration where none of the two strategies is fixed and persistent fluctuations characterize instead the dynamics of the system.

Finally, we show that time-dependent states are observable even in more realistic situations.
Indeed, the assumption that players tend to imitate the strategy played with them by their best neighbors
is only reasonable if such best neighbors are actually over-performing in terms of the payoff.
But why a player who accumulated a payoff better than that of its best neighbor should actually imitate that neighbor's strategy?

To include this possibility, we enriched our model by considering that a given percentage $p$ of randomly selected players does not follow the imitation rule, but rather follows a second {\it introspective} rule which only accounts for the balance between the payoffs accumulated being a cooperator or a defector in the previous step of the game: {$\Delta_{\rm in} g_{i}=(g_i^C-g_i^D)/b$}. In other words, we tested the robustness of our results against variations in the strategy updating rule, replacing in a fraction $p$ of the population the commonly used imitation of the fittest with what sometimes is referred to as myopic strategy updating \cite{amaral2018heterogeneous2}.

Each {\it introspective} player $i$ updates its strategies with all its $k_i$ neighbors, taking as a reference either the strategy $D$, if {$\Delta_{\rm in} g_i\le 0$}, or $C$, if {$\Delta_{\rm in} g_i> 0$}. If player $i$ is already adopting along a given link the reference strategy, the link stays unchanged, but otherwise one has the transition probabilities {$f(C\rightarrow D)= 1/[1+\exp(\Delta_{\rm in} g_i/\alpha)]$ and $f(D\rightarrow C)=1/[1+\exp(-\Delta_{\rm in} g_i/\alpha)]$.
}

Introspective players are destined to assume monotonic identities, as diversity is impossible for them in their asymptotic state. Therefore, it is very relevant to assess how far, in terms of fraction $p$, diversity in the system can survive, and its effects are resilient.
The results are reported in Fig.~\ref{fig4}, and they attest to the fact that the outcomes reported above remain intact to a large extent of heterogeneity in strategy rules. Namely, the overall scenario remains almost unchanged for $p=0.05$ (panel b) and $p=0.1$ (panel c): system-wide shifts in the dominant strategy and time-dependent cooperation are always observed. As the fraction $p$ of introspective players increases up to $p=0.5$ (panel d), one can notice a shrinkage and the eventual disappearance of the full $C$ phase, yet the coexistence of both strategies and thus at least the survival of cooperation remains feasible. Taken together, this rounds up the evidence in favor of the remarkable robustness of the presented results, and it corroborates diverse strategic identities as applicable in a broad plethora of different modeling scenarios.

In summary, we have introduced and studied diverse strategic identities, with a focus on their impact on the resulting evolutionary dynamics. Unlike bacteria and other relatively simple forms of life which use the same strategy for all their interactions, humans and many animals do not. It is indeed straightforward to come up with many examples in support of this claim. From the psychological point of view, differences in personality, lack of confidence, and simply growing up and maturing as an individual are most commonly named as reasons for acting differently around different people \cite{marcia1966development, erikson1994identity}. When studying evolutionary games on networks, it is thus fitting to replace a single strategy of an individual with a strategy vector, such that it is possible for that individual to use different strategies in all its interactions.

We have focused on social dilemmas, and the prisoner's dilemma game in particular, as the most important and frequently used subclass of evolutionary games \cite{rapoport1965prisoner}, and we have considered SF and ER networks of different sizes and with different average degrees. We have also considered part of the network population using an introspective strategy adoption rule besides the more commonly used imitation of the fittest \cite{amaral2018heterogeneous2}. We have shown that, largely regardless of the particularities of the implementation, diverse strategic identities give rise to evolutionary dynamics that is unseen in the monotonic case. Our results include sudden systemic shifts and oscillations of dominant strategies in the network. These sometimes support cooperation where it would otherwise perish, but sometimes also supporting defection where otherwise cooperators would dominate. Furthermore, we have observed time-dependent cooperation and maturation in individual strategic choices, where players initially change their strategies frequently and use different strategies with different neighbors, only to eventually converge to an essentially monotonic strategy in time. However, without the ability to use different strategies with different neighbors, the global evolutionary outcome would have been altogether very different.

Given the high and robust degree to which diverse strategic identities affect the evolutionary dynamics on networks, and given the ubiquity of mathematical models based on evolutionary games in mitigating adverse climate change \cite{vasconcelos_ncc13, pacheco_plrev14}, improving imperfect vaccination outcomes \cite{wang_z_pr16, chen_prsb19}, containing epidemic spreading \cite{pastor_rmp15}, educating against the overuse of antibiotics \cite{chen_fp18}, and preserving biodiversity \cite{szolnoki_jrsif14}, we argue that truly applicable models of the future should relax the {\it one strategy for all} limitation and embrace the freedom and complexity of more realistic diverse strategic identities. As our research shows, this allows individuals to carefully gauge their strategic vectors, which can lead to socially optimal outcomes that would otherwise remain hidden or completely unattainable.

ISN and IL acknowledge
partial support from the Ministerio de Econom\'ia, Industria y Competitividad of Spain under project FIS2017-84151-P.
M.P. acknowledges support by the Slovenian Research Agency (Grant Nos. J4-9302, J1-9112 and P1-0403).
Authors acknowledge the computational resources and assistance provided by CRESCO, the supercomputing center of ENEA in Portici, Italy.

%

\end{document}